\documentclass[useAMS,usenatbib]{mn2e}

\usepackage{times,mathptm}
\usepackage{lscape}
\usepackage{natbib}
\usepackage{graphicx}
\usepackage{floatflt}
\usepackage{amssymb}

\newcommand{\mum}{${\rm \mu m}$}
\newcommand{\mujy}{${\rm \mu Jy}$}
 \newcommand{\ergs}{${\rm ergs~s^{-1}}$}
 \newcommand{\msun}{M$_\odot$}

 \newcommand{\lx}{$L_{\rm X}$}
 \newcommand{\lir}{$L_{\rm IR}$}
 \newcommand{\mgal}{$M_{\ast}$}
 
 \newcommand{\chandra}{{\it Chandra}}
 \newcommand{\herschel}{{\it Herschel}}
 
 \newcommand{\lowz}{low-{\it z}}
 \newcommand{\hiz}{high-{\it z}}
 \newcommand{\avms}{${\rm SFR_{MS}}$}
 \newcommand{\avagn}{$\langle{R_{\rm MS}}\rangle$}
 \newcommand{\rsb}{$R_{\rm MS}$}

 \newcommand{\apjl}{ApJL}
 \newcommand{\apj}{ApJ} 
 \newcommand{\araa}{ARA\&A}
 \newcommand{\aj}{AJ}
 \newcommand{\mnras}{MNRAS}
 
 \newcommand{\apjs}{ApJS}
 
 \newcommand{\aap}{A\&A}

 \title[AGN SFR distributions] {ALMA and \herschel\ reveal that X-ray
   selected AGN and main-sequence galaxies have different star
   formation rate distributions} \author[J. R. Mullaney et al.]
 {J. R. Mullaney$^{1}$\thanks{E-mail: j.mullaney@sheffield.ac.uk},
   D. M. Alexander$^{2}$, J. Aird$^{3}$, E. Bernhard$^{1}$,
   E. Daddi$^{4}$, A. Del Moro$^{2}$, \newauthor M. Dickinson$^{5}$,
   D. Elbaz$^{4}$, C. M. Harrison$^{2}$, S. Juneau$^{4}$,
   D. Liu$^{4}$, M. Pannella$^{6}$,
   \newauthor D. Rosario$^{7}$, P. Santini$^{8}$, M. Sargent$^{9}$, C. Schreiber$^{4}$,  J. Simpson$^{2}$, F. Stanley$^{2}$\\
   $^{1}$Department of Physics and Astronomy, The University of
   Sheffield, Hounsfield
   Road, Sheffield, S3 7RH, UK\\
   $^{2}$Centre of Extragalactic Astronomy, Department of Physics, Durham University, South Road, Durham, DH1 3LE, UK\\
   $^{3}$Institute of Astronomy, University of Cambridge, Madingley
   Road, Cambridge CB3 0HA, UK\\
   $^{4}$Laboratoire AIM, CEA/DSM-CNRS-Universit\'{e} Paris Diderot,
   Irfu/Service d’Astrophysique, CEA-Saclay, Orme des Merisiers, 91191
   Gif-sur-Yvette, France\\
   $^{5}$National Optical Astronomy Observatories, 950 N Cherry Avenue, Tucson, AZ 85719, USA\\
   $^{6}$Universit\"ats-Sternwarte M\"unchen, Scheinerstr. 1, D-81679 M\"unchen\\
   $^{7}$Max-Planck-Institut für Extraterrestrische Physik (MPE), Postfach 1312, 85741, Garching, Germany\\
   $^{8}$INAF-Osservatorio Astronomico di Roma, via di Frascati 33, I-00040 Monte Porzio Catone, Roma, Italy\\
   $^{9}$Astronomy Centre, Department of Physics and Astronomy,
   University of Sussex, Brighton, BN1 9QH, UK}

\begin{document}

\date{Date Accepted}

\maketitle

\begin{abstract} 
  
  Using deep \herschel\ and ALMA observations, we investigate the star
  formation rate (SFR) distributions of X-ray selected AGN host
  galaxies at $0.5<z<1.5$ and $1.5<z<4$, comparing them to that of
  normal, star-forming (i.e., ``main-sequence'', or MS) galaxies.  We
  find 34--55 per cent of AGNs in our sample have SFRs at least a
  factor of two below that of the average MS galaxy, compared to
  $\approx15$ per cent of all MS galaxies, suggesting significantly
  different SFR distributions.  Indeed, when both are modelled as
  log-normal distributions, the mass and redshift-normalised SFR
  distributions of X-ray AGNs are roughly twice as broad, and peak
  $\approx0.4$ dex lower, than that of MS galaxies.  However, like MS
  galaxies, the normalised SFR distribution of AGNs in our sample
  appears not to evolve with redshift.  Despite X-ray AGNs and MS
  galaxies having different SFR distributions, the linear-mean SFR of
  AGNs derived from our distributions is remarkably consistent with
  that of MS galaxies, and thus with previous results derived from
  stacked \herschel\ data.  This apparent contradiction is due to the
  linear-mean SFR being biased by bright outliers, and thus does not
  necessarily represent a true characterisation of the typical SFR of
  X-ray AGNs.
  
\end{abstract}


\section{Introduction}
\label{Introduction}
Today's most successful models of galaxy evolution predict that the
energy released via accretion onto supermassive black holes
(hereafter, BHs) has played an important role in dictating how today's
galaxies have grown and evolved (e.g., \citealt{Schaye15}).  As such,
understanding the connection between galaxy growth via star-formation
and the growth of their resident BHs is one of the key challenges
facing current extragalactic research.  There are now numerous lines
of empirical evidence in support of time-averaged/integrated BH growth
correlating with star-formation in their host galaxies; for example,
(a) the tight proportionality between BH mass and galaxy bulge mass
(e.g., \citealt{Gebhardt00}); (b) the similar cosmic histories of the
volume-averaged BH growth and star formation rates (hereafter, SFR;
e.g., \citealt{Silverman08, Aird15}); and, more directly, (c) the
correlation between average BH growth and SFR among the star-forming
galaxy population (e.g., \citealt{Mullaney12b, Chen13, Delvecchio15,
  Rodighiero15}).  However, it is still far from clear what physical
processes (e.g., feedback processes/common fuel supply/common
triggering mechanism) connect BH growth to star-formation to produce
these average trends.

\begin{figure*}
\begin{center}
  \includegraphics[width=12.3cm]{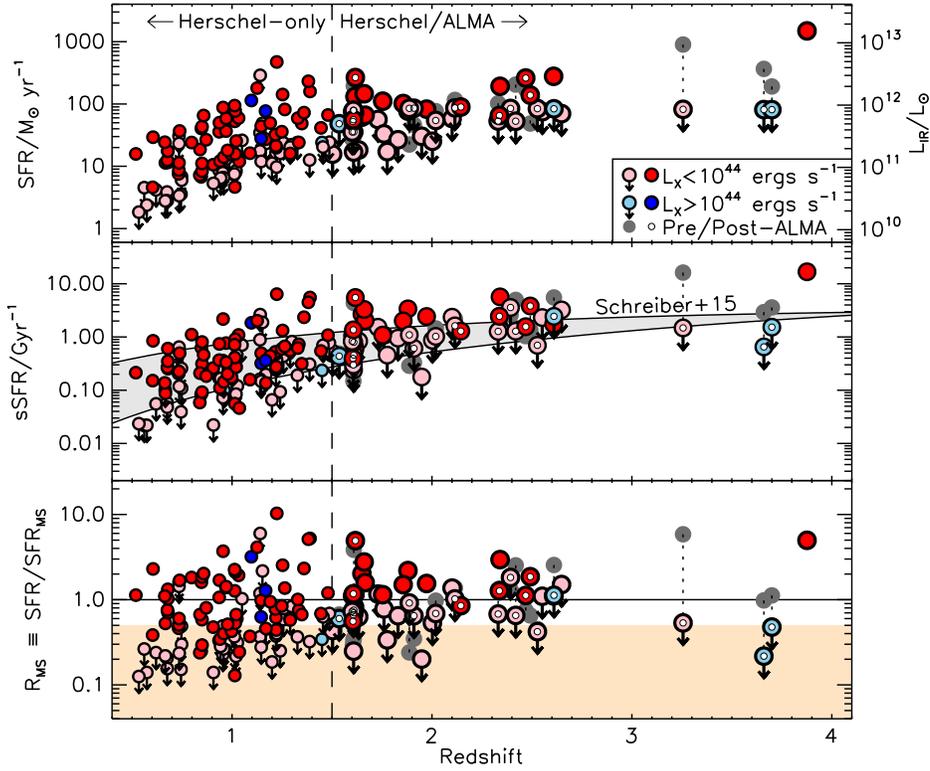}
\end{center}
\caption{Host galaxy star-forming properties of our \lowz\ (i.e.,
  $0.5<z<1.5$; not observed by ALMA) and \hiz\ (i.e., $z>1.5$) samples
  of AGNs (samples separated by the vertical dashed line). In all
  plots, grey circles indicate pre-ALMA (specific) star formation
  rates ($[$s$]$SFRs) from \herschel\, which are connected to their
  ALMA-measured (s)SFRs by dotted lines.  (s)SFRs from ALMA are
  indicated by small white circles.  Red and blue circles represent
  AGNs with \lx$=10^{42-44}$~\ergs\ and \lx$>10^{44}$~\ergs,
  respectively, with lighter colours used for $3\sigma$ upper
  limits. {\it Top:} SFR vs. redshift.  Despite our ALMA observations
  probing SFRs up to a factor of $\approx10$ lower than \herschel,
  only $\approx29$ per cent of our ALMA-targeted AGNs are
  detected. {\it Middle:} sSFR vs. redshift.  In this panel, the
  shaded region represents the average sSFR of main-sequence (MS)
  galaxies (\avms) as described by Eqn. 9 of S15 for the stellar mass
  range of our sample.  {\it Bottom:} \rsb\ vs. redshift.  By
  definition, the horizontal line represents the average \rsb\ of MS
  galaxies.  Shading indicates where \rsb$<0.5$.  Between 34 and 55
  per cent (dependent on upper limits) of AGNs in our combined (i.e.,
  \lowz$+$\hiz) sample lie within this shaded region, compared to
  $\approx15$ per cent of MS galaxies.}
\label{sfr_z}
\end{figure*}

One of the primary means of making progress in this area has been to
measure the SFRs and specific SFRs (i.e., SFR per unit stellar mass,
or sSFR) of galaxies hosting growing BHs (witnessed as active galactic
nuclei, or AGN) and search for correlations or differences (vs. the
non-AGN population) that may signify a causal connection.  The {\it
  Herschel Space Observatory} (hereafter, {\it Herschel}) has played a
major role in progressing this science by providing an
obscuration-independent view of star-formation that is largely
uncontaminated by emission from the AGN.  However, with even the
deepest {\it Herschel} surveys detecting $\lesssim50$ per cent of the
AGN population, most studies have resorted to averaging (often via
stacking analysis, but see \citealt{Stanley15}) to characterise the
(s)SFRs of the AGN population.  These studies have typically reported
that the average SFRs of AGNs trace that of star-forming
``main-sequence'' (hereafter, MS) galaxies (e.g., \citealt{Mullaney12,
  Santini12, Harrison12, Rosario13, Stanley15}), i.e., the dominant
population of star-forming galaxies whose SFRs are roughly
proportional to their stellar mass (i.e., sSFR$\approx$constant), with
a constant of proportionality that increases with redshift (e.g.,
\citealt{Noeske07, Daddi07}).  However, as averages can be biased by
bright outliers, it is feasible that these findings are being driven
upwards by a few bright sources (e.g., Fig. 14 of
\citealt{Rosario15}).  Here, we test this by combining deep \herschel\
and ALMA observations to instead constrain the {\it distribution} of
host galaxy SFRs of a sample of X-ray selected AGNs and comparing it
to that of MS galaxies.  We adopt $H_0 = 71~{\rm km~s^{-1}~Mpc^{-1}}$,
$\Omega_{\Lambda}=0.73$, $\Omega_{\rm M}=0.27$ and a Chabrier initial
mass function (IMF).

\section{Sample selection}
\label{Sample}

To investigate any redshift evolution of the AGN (s)SFR distribution,
we use two samples of X-ray selected AGNs: a \lowz\ sample spanning
$0.5\leq z<1.5$ and a \hiz\ sample spanning $1.5\leq z<4$ (although
the \hiz\ sample is dominated by AGNs at $1.5\leq z<2.7$).  The split
at $z=1.5$ is motivated by our ALMA target selection criteria: for
this, we only consider AGNs with redshifts $>1.5$ since (a) the
majority of $z<1.5$ AGNs are detected with \herschel\ in the deepest
fields and thus already have obscuration-independent SFR measures and
(b) the negative $k$-correction at sub-mm wavelengths would call for
prohibitive ALMA integration times.

The \hiz\ sample were all selected from the 4~Ms \chandra\ Deep Field
South (hereafter, CDF-S) survey catalogue described in \cite{Xue11}
with updated redshifts from \cite{Hsu14}; we recalculate the
rest-frame 2-10~keV luminosities (\lx) of the sources using these new
redshifts.  To ensure reliable AGN selection, we only consider those
sources with \lx$>10^{42}$~\ergs\ and reliable redshifts (spec-$z$, or
phot-$z$ with $\Delta z/(1+z)<0.1$) that lie within 6\arcmin\ of the
average aim point of the survey (the latter ensures reliable positions
for matching to ALMA counterparts).  Our primary science goal of
constraining the SFR distributions of AGN host galaxies in the context
of the MS requires knowledge of the host galaxy stellar masses
(\mgal), which we derive following \cite{Santini12}.  We refer to that
study for a description of the relative uncertainties on \mgal, which
is estimated to be 50 and 20 per cent $(1\sigma)$ for optically
obscured (Type 2) and unobscured (Type 1) AGN, respectively.  Since
the majority (i.e., $>70$ per cent) of the AGNs in our samples are
optically obscured, this level of uncertainty has no significant
impact on our conclusions.  We restrict our sample to AGNs with
\mgal$>2\times10^{10}$~\msun; below this threshold, it becomes
prohibitive to reach low enough flux limits to probe to SFRs
significantly below the mean SFR of MS galaxies (hereafter, \avms)
with ALMA.  Despite this \mgal\ cut we still sample the vast majority
of the luminous AGN population since the \mgal\ distribution of
\lx$>10^{42}$~\ergs\ AGNs peaks at $\approx6\times10^{10}$~\msun\
(e.g., \citealt{Mullaney12}).

\begin{figure}
  \includegraphics[width=8.0cm]{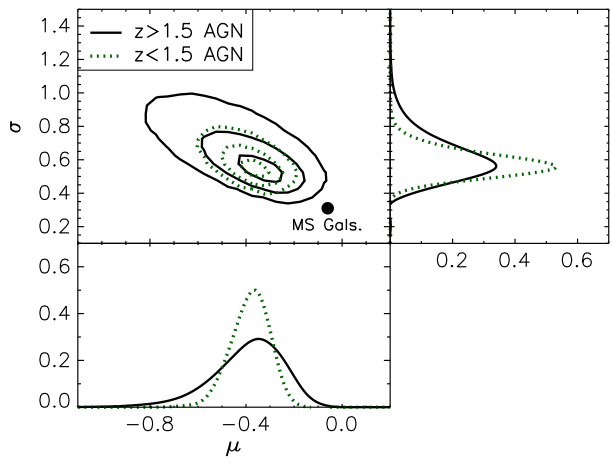}
  \caption{The posterior probability distributions (PDs) for the
    parameters describing the assumed log-normal \rsb\ distribution
    for AGN host galaxies: $\mu$ is the mode of the log-normal, while
    $\sigma$ is its 1$\sigma$ width (see Eq. 1).  PDs for both our
    \lowz\ and \hiz\ samples are shown (see key).  Contours of 20, 68
    and 95 per cent confidence are shown.  The best-fit parameters of
    the combined (i.e., redshift-averaged) \rsb\ distribution of MS
    galaxies is indicated by the solid black circle (from \protect
    \citealt{Schreiber15}). The bottom and rightmost plots indicate
    the relative probability of $\mu$ and $\sigma$ values; the
    location of the peak represent the most probable parameter values.
    When modelled as a log-normal, the \rsb\ distribution of AGN host
    galaxies is significantly broader, and shifted significantly lower
    than that of MS galaxies.}
  \label{par_range}
\end{figure}

The above selection returned 49 AGNs (our \hiz\ sample), with 20 and
29 having spec-$z$ and phot-$z$, respectively.  Of these 49, 13 are
detected in the GOODS-\herschel\ 160~\mum\ maps of the CDF-S
(\citealt{Elbaz11}) from which SFRs are be derived.  Of the remaining
36 AGNs, 24 were observed by ALMA.  However, since making our original
ALMA target list, a more sensitive \herschel\ 160~\mum\ map of the
CDF-S has been generated by combining the PEP (\citealt{Lutz11}) and
GOODS-\herschel\ surveys (\citealt{Magnelli13}) and four of our 24
ALMA targets are now detected in that new map.  For these four, we
adopt the mean (s)SFR derived from the two facilities (see
\S\ref{Obs}).  All other \herschel\ fluxes and $3\sigma$ upper limits
(including for the twelve \herschel -undetected AGNs not targeted by
ALMA) are also taken from the combined PEP$+$GOODS-\herschel\ dataset.

The \lowz\ sample were selected from the regions of the {\it Chandra}
Deep Field North (from \citealt{Alexander03} and adopting the same
redshifts and \mgal\ as \citealt{Mullaney12}) and South
(\citealt{Xue11}, but using the updated redshifts and \mgal) surveys
with {\it Herschel} coverage by the PEP$+$GOODS surveys.  We also
restrict this \lowz\ sample to \lx$>10^{42}$~\ergs\ and \mgal
$>2\times10^{10}$~\msun\ to allow meaningful comparison with the \hiz\
sample.  This returned a sample of 110 AGNs (i.e., our \lowz\ sample),
94 of which have spec-$z$.  Sixty five of these 110 are detected in
the \herschel\ 160~\mum\ band, from which we derive (s)SFRs (see
\S\ref{Obs}); $3\sigma$ flux upper limits were measured for the 45
\herschel\ non-detections.

\begin{table}
  \caption{Best-fit parameters for the
    log-normal \rsb (${\rm=SFR/SFR_{MS}}$) distributions (see Eqn. 1) of the samples of galaxies
    described in the main text.}
\begin{center}
\begin{tabular}{@{}lcc@{}}
\hline
\hline
(1)&(2)&(3)\\
Sample&$\mu$&$\sigma$\\
\hline
MS galaxies (\citealt{Schreiber15})&$-0.06^a$&$0.31\pm0.02$\\
Low-$z$ AGN sample&$-0.378^{+0.068}_{-0.079}$&$0.568^{+0.082}_{-0.062}$\\
High-$z$ AGN sample&$-0.38^{+0.12}_{-0.16}$&$0.59^{+0.15}_{-0.10}$\\
Combined AGN sample&$-0.369^{+0.065}_{-0.080}$&$0.560^{+0.087}_{-0.065}$\\
\hline
\end{tabular}
 
\end{center}{\sc Notes}: Values given are the median of the posterior
probability distributions (PDs) and the 68 per cent confidence
intervals. $^{\rm a}$This is slightly offset from exactly zero as \rsb\
is the SFR relative to the {\it linear mean} SFR of MS galaxies,
whereas $\mu$ is the mode of the \rsb\ distribution.
\label{expec}
\end{table}

\section{ALMA observations and data analysis}
\label{Obs}
All 24 of our ALMA targets were observed with ALMA Band-7 (i.e.,
observed-frame $\sim850$~\mum) during November, 2013, with a longest
baseline of 1.3~km.  To maximise observing efficiency, the
ALMA-targeted sample was split into three groups according to the flux
limit required to probe down to at least \avms\ at a given redshift.
This corresponds to RMS flux limits of 200~\mujy, 125~\mujy\ and
90~\mujy\ for the three groups.  ALMA continuum fluxes were measured
using {\tt uv\_fit} of GILDAS v.apr14c, adopting point source profiles
for two unresolved sources and circular Gaussian profiles for the other
five detected targets.

Measured ALMA and \herschel\ fluxes and upper limits were converted to
8-1000~\mum\ infrared luminosities (hereafter, \lir ) using our
adopted redshifts (see \S2) and the average infrared SEDs of MS
galaxies described in \cite{Bethermin15}, which are constructed using
the theoretical templates of \cite{Draine07}.  However, we note that
our conclusions do not change if we instead use either the
\cite{Chary01} SEDs or a starburst SED (i.e., Arp220).  At the
redshifts of our \hiz\ sample, Band-7 probes the rest-frame
180--340~\mum, close to the peak of the far-infrared emission due to
star-formation.  While these rest-frame wavelengths are also sensitive
to dust mass (e.g., \citealt{Scoville14}), based on the range of
\cite{Draine07} SED templates we estimate that the corresponding \lir\
are accurate to within $\pm0.3$ dex, which we factor into our
analyses.  In a follow-up study we will employ full infrared SED
fitting incorporating all available \herschel\ and ALMA fluxes and
upper limits to reduce the uncertainties associated with the adopted
SED, but such detailed fitting is beyond the scope of this Letter.  As
a check, however, we note that the SFRs derived from ALMA and
\herschel\ data for the four AGNs that are detected with both are
consistent to within this tolerance.  SFRs are derived from \lir\
using Eqn. 4 from \cite{Kennicutt98}, but adopting a Chabrier IMF.
Finally, to explore the distributions of AGN host SFRs relative to
\avms, we define $R_{\rm MS}\equiv {\rm SFR}/{\rm SFR_{MS}}$, the
relative offset from the MS, where \avms\ is computed using Eqn. 9 of
\citeauthor{Schreiber15} (\citeyear{Schreiber15}; hereafter, S15).

\section{Results}
\label{Results}
\subsection{Star-forming properties of X-ray AGNs}
Despite our ALMA observations probing to SFRs up to a factor of
$\approx10$ below that achieved with \herschel\ (Fig. \ref{sfr_z},
top) only seven (i.e., $\approx29$ per cent) of the 24 ALMA-targeted
AGNs in our \hiz\ sample are detected at $>3\sigma$ at 850~\mum.  The
fractions of ALMA-undetected AGN are roughly the same for targets with
spec-$z$ and phot-$z$, suggesting that redshift uncertainties are not
the primary cause of the non-detections.  Despite the high fraction of
non-detections, the $3\sigma$ upper limits provided by the
ALMA$+$\herschel\ data enable us to infer the level of consistency
between the distributions of \rsb\ for AGN and MS galaxies (see
\S\ref{Param}), with the latter having been shown not to vary in the
\mgal\ and redshifts ranges considered here (e.g.,
\citealt{Rodighiero11, Sargent12}).

To explore our AGN hosts' star-forming properties in the context of
the evolving MS, we plot their sSFRs and \rsb\ values as a function of
redshift (Fig.  \ref{sfr_z}, middle and lower panels, respectively).
We find that 54 to 88 (range due to upper limits) of the 159 AGNs
(i.e., $\approx$34 to $\approx$55 per cent) in our combined (i.e.,
\lowz$+$\hiz) sample have \rsb$<0.5$, with significant overlap between
the fractions in our \lowz\ (i.e., $\approx$43 per cent to $\approx$54
per cent) and \hiz\ (i.e., $\approx$14 per cent to $\approx$59 per
cent) samples.  Comparing these fractions to the $\approx$15 per cent
of MS galaxies with \rsb$<0.5$ (from S15), reveals that the AGNs in
our \lowz\ sample, and possibly also our \hiz\ sample, do not trace
the same \rsb\ distribution as MS galaxies, instead displaying a
strong bias toward lower \rsb\ values.  Finally, we note that only
$\approx$5 per cent of AGNs in our combined sample reside in
starbursts (i.e., with \rsb$>4$).

\subsection{Parameterising an X-ray AGN SFR distribution}
\label{Param}
With the large fraction of AGNs with \rsb$<0.5$ in our combined and,
in particular, \lowz\ samples being inconsistent with the \rsb\
distribution of MS galaxies, we now attempt to place constraints on
the distribution of SFRs (relative to the MS; i.e., \rsb) of AGN
hosts.  We place particular emphasis on quantifying the level of
consistency/discrepancy between the AGN and MS \rsb\ distributions.

Our relatively small sample sizes, combined with the large fraction of
non-detections prevents us from determining the AGN \rsb\ distribution
directly.  Since a key goal here is to quantitatively compare the AGN
and MS \rsb\ distributions, we instead {\it assume} the same
log-normal form for the AGN \rsb\ distribution as found for MS
galaxies (e.g., \citealt{Rodighiero11, Sargent12}, S15):

\begin{equation}
  N(R_{\rm MS}) \propto {\rm exp}\left(-\frac{({\rm log}(R_{\rm MS})-\mu)^2}{2\sigma^2}\right)
\end{equation}

\noindent
and infer its parameters (i.e., similar to \citealt{Shao10} who
inferred the AGN \lir\ distribution).  This is done purely to ease
comparison between the AGN and MS \rsb\ distributions by allowing us
to compare like-for-like parameters (i.e., the mode, $\mu$, and the
variance, $\sigma^2$, of the log-normal \rsb\ distribution), and is
not to be taken as a literal description of the true AGN \rsb\
distribution.\footnote{Investigating whether other forms better
  describe the \rsb\ distribution of AGN hosts will be the focus of a
  later study incorporating a larger set of ALMA observations from
  Cycle 2 (PI: Alexander; awaiting completion).}

\begin{figure}
  \includegraphics[width=7.5cm]{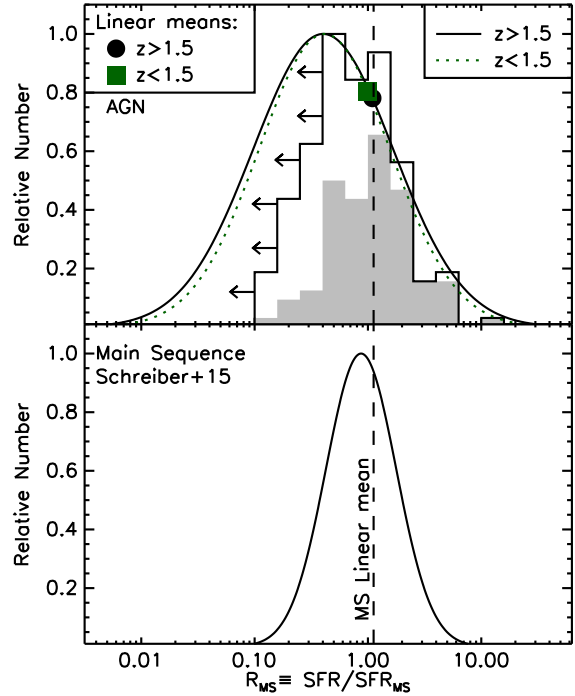}
  \caption{\rsb\ distributions for our \hiz\ and \lowz\ samples of
    X-ray selected AGNs ({\it Top}) and MS galaxies ({\it Bottom};
    from S15).  Here, we show the log-normal distributions with best
    fitting parameters shown in Table 1 (solid and dotted curves; see
    key).  The histograms in the top panel shows the relative numbers
    of AGNs from our combined (i.e., \lowz$+$\hiz) sample in each
    \rsb\ bin; the solid grey histogram represents those AGNs detected
    at $>3\sigma$ with either \herschel\ or ALMA, whereas the empty
    histogram (with left-pointing arrows) also includes upper limits.
    The solid points in the top panel indicate the linear means of the
    log-normal distributions (equivalent to what would be obtained
    via, e.g., stacking analyses) and lie within $1\sigma$ of the
    linear mean \rsb\ of MS galaxies (vertical dashed line).}
\label{ssfrr_dist}
\end{figure}

We adopt a hierarchical Bayesian framework to determine the best-fit
parameters (i.e., $\mu$ and $\sigma$) for our assumed log-normal
distributions, using Gibbs sampling and the Metropolis-Hastings MCMC
algorithm to randomly sample their posterior probability distributions
(hereafter, PDs; \citealt{BDA3}).  The benefits of taking this approach
are that (a) upper limits and uncertainties on \rsb\ can be readily
taken into account and (b) the resulting posterior PDs provide us with
meaningful parameter uncertainties.  We use weak prior PDs, noting
that the centring of these priors (within reasonable limits) has no
significant effect on our results.

The posterior PDs on $\mu$ and $\sigma$ for our two samples are
presented in Fig. \ref{par_range}, while the best-fit parameters
(median of the PDs and 68 per cent confidence intervals) are given in
Table \ref{expec}.  For comparison, we also include the best-fit
parameters of the log-normal \rsb\ distribution for non-AGN MS
galaxies from S15.  As expected from the smaller size of our AGN
sample and the high fractions of non-detections compared to the MS
galaxy sample of S15, the uncertainties on the posterior parameter
values for the assumed AGN log-normal \rsb\ distribution are
considerably larger than those for MS galaxies.  Despite this, our
analysis shows that the \rsb\ distributions of our \lowz\ and \hiz\
AGNs are both significantly broader and peak at significantly lower
values (both at $>99.9$ per cent confidence) than that of MS galaxies.
Interestingly, our analyses show that, as with MS galaxies, there
appears to be little evolution in the AGN \rsb\ distribution, with the
modes and variances of the log-normal distributions describing our
\lowz\ and \hiz\ samples being consistent to within 1$\sigma$.  In
light of this, we infer the \rsb\ distribution of our combined sample,
which we find is roughly twice as broad as, and peaks $\approx0.4$ dex
below, that of MS galaxies (Table \ref{expec}).

\section{Interpretation}

In the previous section we used our combined ALMA+\herschel\ data to
demonstrate that, when modelled as a log-normal, the AGN \rsb\
distribution is significantly broader and peaks at significantly lower
values than that of MS galaxies.  This appears to be at-odds with
recent findings based on mean-stacked \herschel\ data that the average
star-forming properties of AGN hosts is consistent with those of MS
galaxies (e.g., \citealt{Mullaney12, Santini12, Rosario13}).  Here, we
place our results in the context of these studies to explore the root
of these apparent discrepancies.

When comparing to results derived from mean-stacked \herschel\ data,
it is important to note that mean-stacking provides a linear mean
which will not correspond to the mode, $\mu$, of a log-normal
distribution.  Instead, the linear mean will {\it always} be higher
than the mode, with the discrepancy between the two increasing as
function of both $\mu$ and $\sigma$.  Therefore, while results from
mean-stacking still hold when interpreted as the linear mean,
depending on the underlying distribution this may not necessarily
correspond to the mode.

We compare our results against those from stacking by calculating the
linear mean of our log-normal distributions, taking a Monte-Carlo
approach to sample the $\mu$ and $\sigma$ PDs.  This gives linear-mean
AGN \rsb\ values (i.e., \avagn) of $0.99^{+0.23}_{-0.16}$ and
$1.09^{+0.47}_{-0.25}$ for our \lowz\ and \hiz\ samples, respectively
(Fig. \ref{ssfrr_dist}).  These values are remarkably close to the
linear mean \rsb\ of MS galaxies (i.e., \avagn$\approx1$) and are
broadly consistent with the linear means calculated by mean-stacking
\herschel\ 160~\mum\ maps at the positions of our AGN (i.e.,
\avagn$=0.81\pm0.12$ and $0.86\pm0.15$, respectively).  We conclude
that these linear-means are, indeed, influenced by the high tail of
the broad \rsb\ distribution and may not necessarily give a reliable
indication of the modal SFR of AGN hosts.

Despite finding that the \rsb\ distribution of AGN hosts is shifted
toward lower values compared to MS galaxies, our results remain
consistent with AGNs preferentially residing in galaxies with
comparatively high (s)SFRs by $z\sim0$ standards due to the strong
redshift evolution of \avms .  Indeed, applying our analyses to sSFR
(rather than \rsb) gives distributions peaking at $\approx0.2~{\rm
  Gyr^{-1}}$ and $\approx0.5~{\rm Gyr^{-1}}$ for our \lowz\ and \hiz\
samples, respectively.  To put this in context, $\langle{\rm
  sSFR}_{\rm MS}\rangle\approx0.1~{\rm Gyr^{-1}}$ at $z\approx0$, thus
local galaxies with sSFRs of $0.2~{\rm Gyr^{-1}}$ and $0.5~{\rm
  Gyr^{-1}}$ would be classed as MS and starbursting galaxies,
respectively.

Our result compare favourably to those derived from AGN surveys
conducted at other wavelengths.  For example, using SFRs derived from
optical SED fitting, \cite{Bongiorno12} reported a broad sSFR
distribution for X-ray selected AGNs that peaks below that of the MS
at redshifts similar to those explored here (i.e., $0.3<z<2.5$).
Similarly, \cite{Azadi14} showed that the \rsb\ distribution of X-ray
selected AGNs (with a similar \mgal\ selection as here) peaks at
$\sim0.1$ and is similar to the \rsb\ distribution of \mgal-matched
galaxies (i.e., not just star-forming galaxies).  As such, these
studies and the results presented here support the view that X-ray
selected AGN hosts at moderate to high redshifts span the full range
of {\it relative} sSFRs of \mgal$\gtrsim2\times10^{10}$~\msun\
galaxies (e.g., \citealt{Brusa09, Georgakakis14}).  However, with
recent results suggesting that X-ray absorbed AGN may have higher SFRs
than unabsorbed AGN (e.g., \citealt{Juneau13, DelMoro15}), it is
feasible that alternative AGN selections may bring the AGN \rsb\
distribution closer to that of MS galaxies.

\vspace{2mm}
\noindent
We thank the anonymous referee.  DMA, ADM, CMH acknowledge STFC grant
ST/I001573/1. This paper makes use of ALMA data:
ADS/JAO.ALMA\#2012.1.00869.S.
%\bibliography{Mullaney.bib}

\begin{thebibliography}{38}
\expandafter\ifx\csname natexlab\endcsname\relax\def\natexlab#1{#1}\fi

\bibitem[{{Aird} {et~al.}(2015){Aird}, {Coil}, {Georgakakis}, {Nandra},
  {Barro}, \& {P{\'e}rez-Gonz{\'a}lez}}]{Aird15}
{Aird}, J., {Coil}, A.~L., {Georgakakis}, A., {et~al.} 2015, \mnras, 451, 1892

\bibitem[{{Alexander} {et~al.}(2003){Alexander}, {Bauer}, {Brandt},
  {Schneider}, {Hornschemeier}, {Vignali}, {Barger}, {Broos}, {Cowie},
  {Garmire}, {Townsley}, {Bautz}, {Chartas}, \& {Sargent}}]{Alexander03}
{Alexander}, D.~M., {Bauer}, F.~E., {Brandt}, W.~N., {et~al.} 2003, \aj, 126,
  539

\bibitem[{{Azadi} {et~al.}(2014){Azadi}, {Aird}, {Coil}, {Moustakas}, {Mendez},
  {Blanton}, {Cool}, {Eisenstein}, {Wong}, \& {Zhu}}]{Azadi14}
{Azadi}, M., {Aird}, J., {Coil}, A., {et~al.} 2014, ArXiv e-prints, 1407.1975

\bibitem[{{B{\'e}thermin} {et~al.}(2015){B{\'e}thermin}, {Daddi}, {Magdis},
  {Lagos}, {Sargent}, {Albrecht}, {Aussel}, {Bertoldi}, {Buat}, {Galametz},
  {Heinis}, {Ilbert}, {Karim}, {Koekemoer}, {Lacey}, {Le Floc'h}, {Navarrete},
  {Pannella}, {Schreiber}, {Smol{\v c}i{\'c}}, {Symeonidis}, \&
  {Viero}}]{Bethermin15}
{B{\'e}thermin}, M., {Daddi}, E., {Magdis}, G., {et~al.} 2015, \aap, 573, A113

\bibitem[{{Bongiorno} {et~al.}(2012){Bongiorno}, {Merloni}, {Brusa},
  {Magnelli}, {Salvato}, {Mignoli}, {Zamorani}, {Fiore}, {Rosario}, {Mainieri},
  {Hao}, {Comastri}, {Vignali}, {Balestra}, {Bardelli}, {Berta}, {Civano},
  {Kampczyk}, {Le Floc'h}, {Lusso}, {Lutz}, {Pozzetti}, {Pozzi}, {Riguccini},
  {Shankar}, \& {Silverman}}]{Bongiorno12}
{Bongiorno}, A., {Merloni}, A., {Brusa}, M., {et~al.} 2012, \mnras, 427, 3103

\bibitem[{{Brusa} {et~al.}(2009){Brusa}, {Fiore}, {Santini}, {Grazian},
  {Comastri}, {Zamorani}, {Hasinger}, {Merloni}, {Civano}, {Fontana}, \&
  {Mainieri}}]{Brusa09}
{Brusa}, M., {Fiore}, F., {Santini}, P., {et~al.} 2009, \aap, 507, 1277

\bibitem[{{Chary} \& {Elbaz}(2001)}]{Chary01}
{Chary}, R., \& {Elbaz}, D. 2001, \apj, 556, 562

\bibitem[{{Chen} {et~al.}(2013){Chen}, {Hickox}, {Alberts}, {Brodwin}, {Jones},
  {Murray}, {Alexander}, {Assef}, {Brown}, {Dey}, {Forman}, {Gorjian},
  {Goulding}, {Le Floc'h}, {Jannuzi}, {Mullaney}, \& {Pope}}]{Chen13}
{Chen}, C.-T.~J., {Hickox}, R.~C., {Alberts}, S., {et~al.} 2013, \apj, 773, 3

\bibitem[{{Daddi} {et~al.}(2007){Daddi}, {Dickinson}, {Morrison}, {Chary},
  {Cimatti}, {Elbaz}, {Frayer}, {Renzini}, {Pope}, {Alexander}, {Bauer},
  {Giavalisco}, {Huynh}, {Kurk}, \& {Mignoli}}]{Daddi07}
{Daddi}, E., {Dickinson}, M., {Morrison}, G., {et~al.} 2007, \apj, 670, 156

\bibitem[{{Del Moro} {et~al.}(2015){Del Moro}, {Alexander}, {Bauer}, {Daddi},
  {Kocevski}, {McIntosh}, {Brandt}, {Elbaz}, {Harrison}, {Luo}, {Mullaney},
  {Stanley}, \& {Xue}}]{DelMoro15}
{Del Moro}, A., {Alexander}, D.~M., {Bauer}, F.~E., {et~al.} 2015, ArXiv
  e-prints

\bibitem[{{Delvecchio} {et~al.}(2015){Delvecchio}, {Lutz}, {Berta}, {Rosario},
  {Zamorani}, {Pozzi}, {Gruppioni}, {Vignali}, {Brusa}, {Cimatti}, {Clements},
  {Cooray}, {Farrah}, {Lanzuisi}, {Oliver}, {Rodighiero}, {Santini}, \&
  {Symeonidis}}]{Delvecchio15}
{Delvecchio}, I., {Lutz}, D., {Berta}, S., {et~al.} 2015, ArXiv e-prints,
  1501.07602

\bibitem[{{Draine} \& {Li}(2007)}]{Draine07}
{Draine}, B.~T., \& {Li}, A. 2007, \apj, 657, 810

\bibitem[{{Elbaz} {et~al.}(2011){Elbaz}, {Dickinson}, {Hwang},
  {D{\'{\i}}az-Santos}, {Magdis}, {Magnelli}, {Le Borgne}, {Galliano},
  {Pannella}, {Chanial}, {Armus}, {Charmandaris}, {Daddi}, {Aussel}, {Popesso},
  {Kartaltepe}, {Altieri}, {Valtchanov}, {Coia}, {Dannerbauer}, {Dasyra},
  {Leiton}, {Mazzarella}, {Alexander}, {Buat}, {Burgarella}, {Chary}, {Gilli},
  {Ivison}, {Juneau}, {Le Floc'h}, {Lutz}, {Morrison}, {Mullaney}, {Murphy},
  {Pope}, {Scott}, {Brodwin}, {Calzetti}, {Cesarsky}, {Charlot}, {Dole},
  {Eisenhardt}, {Ferguson}, {F{\"o}rster Schreiber}, {Frayer}, {Giavalisco},
  {Huynh}, {Koekemoer}, {Papovich}, {Reddy}, {Surace}, {Teplitz}, {Yun}, \&
  {Wilson}}]{Elbaz11}
{Elbaz}, D., {Dickinson}, M., {Hwang}, H.~S., {et~al.} 2011, \aap, 533, A119

\bibitem[{{Gebhardt} {et~al.}(2000){Gebhardt}, {Bender}, {Bower}, {Dressler},
  {Faber}, {Filippenko}, {Green}, {Grillmair}, {Ho}, {Kormendy}, {Lauer},
  {Magorrian}, {Pinkney}, {Richstone}, \& {Tremaine}}]{Gebhardt00}
{Gebhardt}, K., {Bender}, R., {Bower}, G., {et~al.} 2000, \apjl, 539, L13

\bibitem[{{Gelman} {et~al.}(2014){Gelman}, {Carlin}, {Stern}, {Dunson},
  {Vehtari}, \& {Rubin}}]{BDA3}
{Gelman}, A., {Carlin}, J.~B., {Stern}, H.~S., {et~al.} 2014, {Bayesian Data
  Analysis, 3rd.~ed}

\bibitem[{{Georgakakis} {et~al.}(2014){Georgakakis}, {P{\'e}rez-Gonz{\'a}lez},
  {Fanidakis}, {Salvato}, {Aird}, {Messias}, {Lotz}, {Barro}, {Hsu}, {Nandra},
  {Rosario}, {Cooper}, {Kocevski}, \& {Newman}}]{Georgakakis14}
{Georgakakis}, A., {P{\'e}rez-Gonz{\'a}lez}, P.~G., {Fanidakis}, N., {et~al.}
  2014, \mnras, 440, 339

\bibitem[{{Harrison} {et~al.}(2012){Harrison}, {Alexander}, {Mullaney},
  {Altieri}, {Coia}, {Charmandaris}, {Daddi}, {Dannerbauer}, {Dasyra}, {Del
  Moro}, {Dickinson}, {Hickox}, {Ivison}, {Kartaltepe}, {Le Floc'h}, {Leiton},
  {Magnelli}, {Popesso}, {Rovilos}, {Rosario}, \& {Swinbank}}]{Harrison12}
{Harrison}, C.~M., {Alexander}, D.~M., {Mullaney}, J.~R., {et~al.} 2012, \apjl,
  760, L15

\bibitem[{{Hsu} {et~al.}(2014){Hsu}, {Salvato}, {Nandra}, {Brusa}, {Bender},
  {Buchner}, {Donley}, {Kocevski}, {Guo}, {Hathi}, {Rangel}, {Willner},
  {Brightman}, {Georgakakis}, {Budav{\'a}ri}, {Szalay}, {Ashby}, {Barro},
  {Dahlen}, {Faber}, {Ferguson}, {Galametz}, {Grazian}, {Grogin}, {Huang},
  {Koekemoer}, {Lucas}, {McGrath}, {Mobasher}, {Peth}, {Rosario}, \&
  {Trump}}]{Hsu14}
{Hsu}, L.-T., {Salvato}, M., {Nandra}, K., {et~al.} 2014, \apj, 796, 60

\bibitem[{{Juneau} {et~al.}(2013){Juneau}, {Dickinson}, {Bournaud},
  {Alexander}, {Daddi}, {Mullaney}, {Magnelli}, {Kartaltepe}, {Hwang},
  {Willner}, {Coil}, {Rosario}, {Trump}, {Weiner}, {Willmer}, {Cooper},
  {Elbaz}, {Faber}, {Frayer}, {Kocevski}, {Laird}, {Monkiewicz}, {Nandra},
  {Newman}, {Salim}, \& {Symeonidis}}]{Juneau13}
{Juneau}, S., {Dickinson}, M., {Bournaud}, F., {et~al.} 2013, \apj, 764, 176

\bibitem[{{Kennicutt}(1998)}]{Kennicutt98}
{Kennicutt}, Jr., R.~C. 1998, \araa, 36, 189

\bibitem[{{Lutz} {et~al.}(2011){Lutz}, {Poglitsch}, {Altieri}, {Andreani},
  {Aussel}, {Berta}, {Bongiovanni}, {Brisbin}, {Cava}, {Cepa}, {Cimatti},
  {Daddi}, {Dominguez-Sanchez}, {Elbaz}, {F{\"o}rster Schreiber}, {Genzel},
  {Grazian}, {Gruppioni}, {Harwit}, {Le Floc'h}, {Magdis}, {Magnelli},
  {Maiolino}, {Nordon}, {P{\'e}rez Garc{\'{\i}}a}, {Popesso}, {Pozzi},
  {Riguccini}, {Rodighiero}, {Saintonge}, {Sanchez Portal}, {Santini}, {Shao},
  {Sturm}, {Tacconi}, {Valtchanov}, {Wetzstein}, \& {Wieprecht}}]{Lutz11}
{Lutz}, D., {Poglitsch}, A., {Altieri}, B., {et~al.} 2011, \aap, 532, A90

\bibitem[{{Magnelli} {et~al.}(2013){Magnelli}, {Popesso}, {Berta}, {Pozzi},
  {Elbaz}, {Lutz}, {Dickinson}, {Altieri}, {Andreani}, {Aussel},
  {B{\'e}thermin}, {Bongiovanni}, {Cepa}, {Charmandaris}, {Chary}, {Cimatti},
  {Daddi}, {F{\"o}rster Schreiber}, {Genzel}, {Gruppioni}, {Harwit}, {Hwang},
  {Ivison}, {Magdis}, {Maiolino}, {Murphy}, {Nordon}, {Pannella}, {P{\'e}rez
  Garc{\'{\i}}a}, {Poglitsch}, {Rosario}, {Sanchez-Portal}, {Santini}, {Scott},
  {Sturm}, {Tacconi}, \& {Valtchanov}}]{Magnelli13}
{Magnelli}, B., {Popesso}, P., {Berta}, S., {et~al.} 2013, \aap, 553, A132

\bibitem[{{Mullaney} {et~al.}(2012{\natexlab{a}}){Mullaney}, {Pannella},
  {Daddi}, {Alexander}, {Elbaz}, {Hickox}, {Bournaud}, {Altieri}, {Aussel},
  {Coia}, {Dannerbauer}, {Dasyra}, {Dickinson}, {Hwang}, {Kartaltepe},
  {Leiton}, {Magdis}, {Magnelli}, {Popesso}, {Valtchanov}, {Bauer}, {Brandt},
  {Del Moro}, {Hanish}, {Ivison}, {Juneau}, {Luo}, {Lutz}, {Sargent}, {Scott},
  \& {Xue}}]{Mullaney12}
{Mullaney}, J.~R., {Pannella}, M., {Daddi}, E., {et~al.} 2012{\natexlab{a}},
  \mnras, 419, 95

\bibitem[{{Mullaney} {et~al.}(2012{\natexlab{b}}){Mullaney}, {Daddi},
  {B{\'e}thermin}, {Elbaz}, {Juneau}, {Pannella}, {Sargent}, {Alexander}, \&
  {Hickox}}]{Mullaney12b}
{Mullaney}, J.~R., {Daddi}, E., {B{\'e}thermin}, M., {et~al.}
  2012{\natexlab{b}}, \apjl, 753, L30

\bibitem[{{Noeske} {et~al.}(2007){Noeske}, {Weiner}, {Faber}, {Papovich},
  {Koo}, {Somerville}, {Bundy}, {Conselice}, {Newman}, {Schiminovich}, {Le
  Floc'h}, {Coil}, {Rieke}, {Lotz}, {Primack}, {Barmby}, {Cooper}, {Davis},
  {Ellis}, {Fazio}, {Guhathakurta}, {Huang}, {Kassin}, {Martin}, {Phillips},
  {Rich}, {Small}, {Willmer}, \& {Wilson}}]{Noeske07}
{Noeske}, K.~G., {Weiner}, B.~J., {Faber}, S.~M., {et~al.} 2007, \apjl, 660,
  L43

\bibitem[{{Rodighiero} {et~al.}(2011){Rodighiero}, {Daddi}, {Baronchelli},
  {Cimatti}, {Renzini}, {Aussel}, {Popesso}, {Lutz}, {Andreani}, {Berta},
  {Cava}, {Elbaz}, {Feltre}, {Fontana}, {F{\"o}rster Schreiber},
  {Franceschini}, {Genzel}, {Grazian}, {Gruppioni}, {Ilbert}, {Le Floch},
  {Magdis}, {Magliocchetti}, {Magnelli}, {Maiolino}, {McCracken}, {Nordon},
  {Poglitsch}, {Santini}, {Pozzi}, {Riguccini}, {Tacconi}, {Wuyts}, \&
  {Zamorani}}]{Rodighiero11}
{Rodighiero}, G., {Daddi}, E., {Baronchelli}, I., {et~al.} 2011, \apjl, 739,
  L40

\bibitem[{{Rodighiero} {et~al.}(2015){Rodighiero}, {Brusa}, {Daddi},
  {Negrello}, {Mullaney}, {Delvecchio}, {Lutz}, {Renzini}, {Franceschini},
  {Baronchelli}, {Pozzi}, {Gruppioni}, {Strazzullo}, {Cimatti}, \&
  {Silverman}}]{Rodighiero15}
{Rodighiero}, G., {Brusa}, M., {Daddi}, E., {et~al.} 2015, \apjl, 800, L10

\bibitem[{{Rosario} {et~al.}(2013){Rosario}, {Trakhtenbrot}, {Lutz}, {Netzer},
  {Trump}, {Silverman}, {Schramm}, {Lusso}, {Berta}, {Bongiorno}, {Brusa},
  {F{\"o}rster-Schreiber}, {Genzel}, {Lilly}, {Magnelli}, {Mainieri},
  {Maiolino}, {Merloni}, {Mignoli}, {Nordon}, {Popesso}, {Salvato}, {Santini},
  {Tacconi}, \& {Zamorani}}]{Rosario13}
{Rosario}, D.~J., {Trakhtenbrot}, B., {Lutz}, D., {et~al.} 2013, \aap, 560, A72

\bibitem[{{Rosario} {et~al.}(2015){Rosario}, {McIntosh}, {van der Wel},
  {Kartaltepe}, {Lang}, {Santini}, {Wuyts}, {Lutz}, {Rafelski}, {Villforth},
  {Alexander}, {Bauer}, {Bell}, {Berta}, {Brandt}, {Conselice}, {Dekel},
  {Faber}, {Ferguson}, {Genzel}, {Grogin}, {Kocevski}, {Koekemoer}, {Koo},
  {Lotz}, {Magnelli}, {Maiolino}, {Mozena}, {Mullaney}, {Papovich}, {Popesso},
  {Tacconi}, {Trump}, {Avadhuta}, {Bassett}, {Bell}, {Bernyk}, {Bournaud},
  {Cassata}, {Cheung}, {Croton}, {Donley}, {DeGroot}, {Guedes}, {Hathi},
  {Herrington}, {Hilton}, {Lai}, {Lani}, {Martig}, {McGrath}, {Mutch},
  {Mortlock}, {McPartland}, {O'Leary}, {Peth}, {Pillepich}, {Poole}, {Snyder},
  {Straughn}, {Telford}, {Tonini}, \& {Wandro}}]{Rosario15}
{Rosario}, D.~J., {McIntosh}, D.~H., {van der Wel}, A., {et~al.} 2015, \aap,
  573, A85

\bibitem[{{Santini} {et~al.}(2012){Santini}, {Rosario}, {Shao}, {Lutz},
  {Maiolino}, {Alexander}, {Altieri}, {Andreani}, {Aussel}, {Bauer}, {Berta},
  {Bongiovanni}, {Brandt}, {Brusa}, {Cepa}, {Cimatti}, {Daddi}, {Elbaz},
  {Fontana}, {F{\"o}rster Schreiber}, {Genzel}, {Grazian}, {Le Floc'h},
  {Magnelli}, {Mainieri}, {Nordon}, {P{\'e}rez Garcia}, {Poglitsch}, {Popesso},
  {Pozzi}, {Riguccini}, {Rodighiero}, {Salvato}, {Sanchez-Portal}, {Sturm},
  {Tacconi}, {Valtchanov}, \& {Wuyts}}]{Santini12}
{Santini}, P., {Rosario}, D.~J., {Shao}, L., {et~al.} 2012, \aap, 540, A109

\bibitem[{{Sargent} {et~al.}(2012){Sargent}, {B{\'e}thermin}, {Daddi}, \&
  {Elbaz}}]{Sargent12}
{Sargent}, M.~T., {B{\'e}thermin}, M., {Daddi}, E., \& {Elbaz}, D. 2012, \apjl,
  747, L31

\bibitem[{{Schaye} {et~al.}(2015){Schaye}, {Crain}, {Bower}, {Furlong},
  {Schaller}, {Theuns}, {Dalla Vecchia}, {Frenk}, {McCarthy}, {Helly},
  {Jenkins}, {Rosas-Guevara}, {White}, {Baes}, {Booth}, {Camps}, {Navarro},
  {Qu}, {Rahmati}, {Sawala}, {Thomas}, \& {Trayford}}]{Schaye15}
{Schaye}, J., {Crain}, R.~A., {Bower}, R.~G., {et~al.} 2015, \mnras, 446, 521

\bibitem[{{Schreiber} {et~al.}(2015){Schreiber}, {Pannella}, {Elbaz},
  {B{\'e}thermin}, {Inami}, {Dickinson}, {Magnelli}, {Wang}, {Aussel}, {Daddi},
  {Juneau}, {Shu}, {Sargent}, {Buat}, {Faber}, {Ferguson}, {Giavalisco},
  {Koekemoer}, {Magdis}, {Morrison}, {Papovich}, {Santini}, \&
  {Scott}}]{Schreiber15}
{Schreiber}, C., {Pannella}, M., {Elbaz}, D., {et~al.} 2015, \aap, 575, A74

\bibitem[{{Scoville} {et~al.}(2014){Scoville}, {Aussel}, {Sheth}, {Scott},
  {Sanders}, {Ivison}, {Pope}, {Capak}, {Vanden Bout}, {Manohar}, {Kartaltepe},
  {Robertson}, \& {Lilly}}]{Scoville14}
{Scoville}, N., {Aussel}, H., {Sheth}, K., {et~al.} 2014, \apj, 783, 84

\bibitem[{{Shao} {et~al.}(2010){Shao}, {Lutz}, {Nordon}, {Maiolino},
  {Alexander}, {Altieri}, {Andreani}, {Aussel}, {Bauer}, {Berta},
  {Bongiovanni}, {Brandt}, {Brusa}, {Cava}, {Cepa}, {Cimatti}, {Daddi},
  {Dominguez-Sanchez}, {Elbaz}, {F{\"o}rster Schreiber}, {Geis}, {Genzel},
  {Grazian}, {Gruppioni}, {Magdis}, {Magnelli}, {Mainieri}, {P{\'e}rez
  Garc{\'{\i}}a}, {Poglitsch}, {Popesso}, {Pozzi}, {Riguccini}, {Rodighiero},
  {Rovilos}, {Saintonge}, {Salvato}, {Sanchez Portal}, {Santini}, {Sturm},
  {Tacconi}, {Valtchanov}, {Wetzstein}, \& {Wieprecht}}]{Shao10}
{Shao}, L., {Lutz}, D., {Nordon}, R., {et~al.} 2010, \aap, 518, L26

\bibitem[{{Silverman} {et~al.}(2008){Silverman}, {Green}, {Barkhouse}, {Kim},
  {Kim}, {Wilkes}, {Cameron}, {Hasinger}, {Jannuzi}, {Smith}, {Smith}, \&
  {Tananbaum}}]{Silverman08}
{Silverman}, J.~D., {Green}, P.~J., {Barkhouse}, W.~A., {et~al.} 2008, \apj,
  679, 118

\bibitem[{{Stanley} {et~al.}(2015){Stanley}, {Harrison}, {Alexander},
  {Swinbank}, {Aird}, {Del Moro}, {Hickox}, \& {Mullaney}}]{Stanley15}
{Stanley}, F., {Harrison}, C.~M., {Alexander}, D.~M., {et~al.} 2015, ArXiv
  e-prints, 1502.07756

\bibitem[{{Xue} {et~al.}(2011){Xue}, {Luo}, {Brandt}, {Bauer}, {Lehmer},
  {Broos}, {Schneider}, {Alexander}, {Brusa}, {Comastri}, {Fabian}, {Gilli},
  {Hasinger}, {Hornschemeier}, {Koekemoer}, {Liu}, {Mainieri}, {Paolillo},
  {Rafferty}, {Rosati}, {Shemmer}, {Silverman}, {Smail}, {Tozzi}, \&
  {Vignali}}]{Xue11}
{Xue}, Y.~Q., {Luo}, B., {Brandt}, W.~N., {et~al.} 2011, \apjs, 195, 10

\end{thebibliography}

\end{document}